\documentclass[letter, traditabstract]{aa} % for the letters 
\usepackage{epsfig}
\usepackage{wasysym}
\usepackage{graphicx}
\usepackage{natbib}
\usepackage{lscape}

\usepackage{color}
\providecommand{\tabularnewline}{\\}
\usepackage{txfonts}
\begin{document}

\title{On the origin of stars with and without planets.}

\subtitle{T$_{c}$ trends and clues to Galactic evolution \thanks{
Based on observations collected with the HARPS spectrograph
at the 3.6-m telescope (072.C-0488(E)), installed at the La Silla
Observatory, ESO (Chile), with the UVES spectrograph at the 8-m
Very Large Telescope  – program IDs: 67.C-0206(A), 074.C-
0134(A), 075.D-0453(A) –, installed at the Cerro Paranal Observatory,
ESO (Chile), and with the UES spectrograph at the 4.2-m William
Herschel Telescope, installed at the Spanish Observatorio del
Roque de los Muchachos of the Instituto de Astrofísica de Canarias, on
the island of La Palma.}}

\titlerunning{Tc slope}

%\thanks{}

\author{V.~Zh.~Adibekyan\inst{1} 
\and J.~I.~Gonz\'{a}lez Hern\'{a}ndez \inst{2,3}
\and E.~Delgado~Mena\inst{1}
\and S.~G.~Sousa\inst{1,2,4}
\and \\ N.~C.~Santos\inst{1,4}
\and G.~Israelian\inst{2,3}
\and P.~ Figueira\inst{1}
\and S. Bertran de Lis\inst{2,3}
}

\institute{Centro de Astrof\'{\i}ísica da Universidade do Porto, Rua das Estrelas,
4150-762 Porto, Portugal\\
\email{Vardan.Adibekyan@astro.up.pt}
\and Instituto de Astrof\'{\i}sica de Canarias, 38200 La Laguna, Tenerife, Spain
\and Departamento de Astrof{\'\i}sica, Universidad de La Laguna, 38206 La Laguna, Tenerife, Spain
\and Departamento de F\'{\i}ísica e Astronomia, Faculdade de Ci\^{e}ncias da Universidade do Porto, Portugal}

   \date{Received ... / Accepted ...}

% \abstract{}{}{}{}{} 
% 5 {} token are mandatory
 
\abstract
{We explore a sample of 148 solar-like stars to search for  a possible correlation between the slopes of the abundance trends versus 
condensation temperature (known as the T$_{c}$ slope) with stellar parameters and Galactic orbital parameters in order to understand the nature of 
the peculiar chemical signatures of these stars and the possible connection with planet formation. 
We find that the T$_{c}$ slope significantly correlates (at more than 4$\sigma$)   
with the stellar age and the stellar surface gravity. We also find tentative evidence that the T$_{c}$ slope correlates with the mean 
galactocentric distance of 
the stars (R$_{mean}$), suggesting that those stars that originated in the inner Galaxy have fewer refractory elements relative to the volatiles.
While the average T$_{c}$ slope for planet-hosting solar analogs is steeper than that of their counterparts without planets, this difference
probably reflects the difference in their age and R$_{mean}$.
We conclude that the age and probably the Galactic birth place are determinant to establish the star's chemical properties. 
Old stars (and stars with
inner disk origin) have a lower refractory-to-volatile ratio.}


\maketitle

%
%________________________________________________________________________________________________________

\section{Introduction}

Despite the huge progress in developing instrumentation and observational techniques during the past decade,
the study of extrasolar planets' properties via direct observations is still a very difficult task, and
the precise study and characterization of known extroplanets cannot be dissociated from the study of planet host stars.

The connection between stellar and planetary properties has been widely explored. In particular, the very first correlation 
observed in this field of research, that of giant-planet  - metallicity \citep[e.g.,][]{Gonzalez-97, Santos-01, Santos-04, Fischer-05, Sousa-11}, was one of 
the most important constraints on planet formation theories \citep[e.g.,][]{Mordasini-09}. Afterwards it was shown that
not only does the presence of planets correlate with metallicity (usually abundance of iron), but planet-host stars also show a systematic 
enhancement of other elements \citep[e.g., $\alpha$-elements - ][]{Haywood-08b, Adibekyan-12a, Adibekyan-12b}. More recently, several studies
have revealed that 
stellar metallicity also plays an important role in the architecture of planetary orbits \citep[e.g.,][]{Dawson-13, Beauge-13, Adibekyan-13a}.

Naturally, this connection is bidirectional: not only do stellar properties play an important role in planet formation 
and evolution, but the planet formation can also have an impact on stellar properties. As an example, \cite{Israelian-09} and \cite{DelgadoMena-13}
find strong evidence that solar analogs with planets show enhanced Li depletion compared to their non-host counterparts.
The studies mentioned above are not the only ones.  Recently, \cite{Melendez-09} have claimed that the Sun shows a deficiency in refractory 
elements with respect to other solar twins probably because they were trapped in the terrestrial planets in our solar system. 
The same conclusion has also been reached by \cite{Ramirez-09}  who analyzed 64 solar twins and analogs.  However, recent results by 
\cite[][- GH10,13]{Jonay-10, Jonay-13} strongly challenge the relation between the presence of planets and the abundance peculiarities of the stars.
Other works also have examined this possible connection between the chemical peculiarities and formation of planets 
\citep[e.g.,][]{Smith-01, Ecuvillon-06, Gonzalez-10, Gonzalez-11, Jonay-11, Schuler-11, Ramirez-14}.

In this letter, we explore the origin of the possible trend observed between [X/H] (or [X/Fe]) and condensation 
temperature ($T_{c}$) using a sample of 148 solar-like stars.

\section{Data}

Our initial sample is a combination of two samples of solar analogs (95 stars) and ``hot'' analogs (61 stars) taken from 
GH10,13. We have
cross-matched this sample with the Geneva-Copenhagen Survey sample \citep[GCS-][]{Nordstrom-04}, for which \cite{Casagrande-11}
provides the Galactic orbital parameters, the space velocity components, and the ages of 148 of the stars considered in our study%
\footnote{Throughout the paper, BASTI expectation ages are used as suggested by \cite{Casagrande-11}.%
}. Fifty-seven of these stars are planet hosts, while for the remaining 91 no planetary companion has been detected up to now.

The stellar atmospheric parameters and the slopes of the $\Delta$[X/Fe]$_{SUN-star}$ versus  T$_{c}$ were derived using very high-quality 
HARPS spectra%
\footnote{Zero slope means solar chemical composition, and a positive slope corresponds to a smaller refractory-to-volatile ratio compared to the Sun.%
}. Twenty-five elements from C (Z = 6) to Eu (Z = 63) have been used for this analysis. 
These slopes are corrected for the Galactic chemical evolution trends as discussed in GH10,13.  

The stars in the sample have effective temperatures 5604 \emph{K} $\leq$ T$_{eff}$ $\leq$ 6374 \emph{K}, 
metallicites -0.29 $\leq$ {[}Fe/H{]} $\leq$ 0.38 dex, and surface gravities
4.14 $\leq$ $\log g$ $\leq$ 4.63 dex. Throughout the letter we defined solar analogs as stars with; T$_{eff}$ = 5777$\pm$200 K; 
logg = 4.44$\pm$0.20 dex; [Fe/H] = 0.0$\pm$0.2 dex. Solar twins are defined as with; T$_{eff}$ = 5777$\pm$100 K; 
logg = 4.44$\pm$0.10 dex; [Fe/H] = 0.0$\pm$0.1 dex. Fifteen out of 58 solar analogs in this sample are known to be orbited by planets and
three out of 15 solar twins are planet hosts.

We would like to point out that stellar parameters derived in GH10,13 and in \cite{Casagrande-11} are in perfect agreement
($\Delta$T$_{eff}$ = -8$\pm$50 K, $\Delta$ $\log g$ = 0.06$\pm$0.07 dex, and $\Delta$[Fe/H] = -0.02$\pm$0.06 dex), which means
that the ages derived in \cite{Casagrande-11} are indeed coherent with the rest of the observables.

\begin{table}
\centering
\caption{Significance levels of the obtained correlations.}
\label{ks}
\begin{tabular}{lcccc}
\hline
\hline 
Parameter & Cor. coeff & Significance & Cor. coeff & Significance \tabularnewline
& \multicolumn{2}{c}{  All } & \multicolumn{2}{c}{Solar analogs}\tabularnewline
\hline 
Age  & 0.35 & \textbf{4.37} & 0.60 & \textbf{4.57}\tabularnewline
$\log g$  & -0.50 & \textbf{6.01} & -0.73 & \textbf{5.57}\tabularnewline
$[$Fe/H$]$  & -0.14 & 1.73 & 0.00 & 0.05\tabularnewline
T$_{eff}$ & -0.05 & 0.59 & 0.03 & 0.17\tabularnewline

\hline 
\end{tabular}
\end{table}

\section{Correlations with Tc slope}

We searched for possible correlations between the T$_{c}$ slope and, in turn, atmospheric parameters and also Galactic orbital parameters and age,
in order to understand which is/are the main factor(s) responsible for the abundance trends with T$_{c}$.

\subsection{ T$_{c}$ slope against stellar parameters and age}

In Fig.~\ref{fig_slope_logg} we show the relation between the T$_{c}$ slopes and $\log g$  both for the full sample and for the
solar analogs exclusively. One can note that the T$_{c}$ slopes strongly correlate with the $\log g$. To evaluate the significance of the correlation
we performed a simple bootstrapped Monte Carlo test. For more details about the test we refer the reader to \cite{Figueira-13}
and \cite{Adibekyan-13b}.
The correlation coefficients and the significance levels (z-scores) of the correlations are presented in Table 1. The same table 
shows that T$_{c}$ slopes do not significantly correlate with other stellar parameters. The T$_{c}$ slope versus {[}Fe/H{]} and T$_{eff}$ plots 
are shown in Fig.~\ref{fig_slope_atmos_param}.

\begin{figure}
\begin{center}
\begin{tabular}{c}
\includegraphics[angle=0,width=0.8\linewidth]{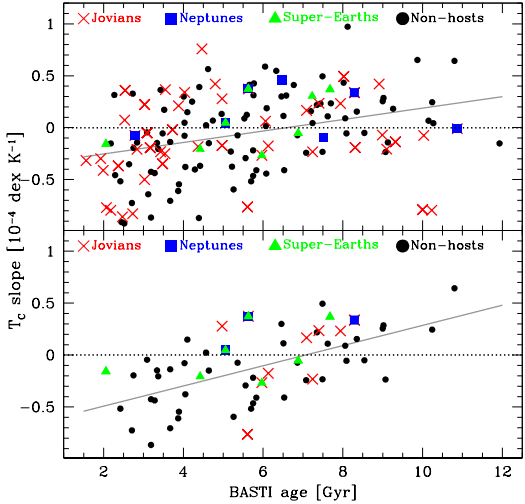}
\end{tabular}
\end{center}
\vspace{-0.6cm}
\caption{T$_{c}$ slopes versus ages for the full sample (\textit{top}) and for the solar analogs (\textit{bottom}).
Gray solid lines provide linear fits to the data points.}
\label{fig_slope_age}
\end{figure}

The observed significant relation between the surface gravity and T$_{c}$ slope means that the chemical abundance trends depend
either on the evolutionary stage of the star or on its age%
\footnote{Since a solar type star evolves from ZAMS, its $\log g$  slowly increases with age along the main sequence.%
}. In Fig.~\ref{fig_slope_age} we plot the T$_{c}$ slope against
the stellar age. The plot and Table 1 clearly show that the correlation with age is quite significant and confirms the result obtained for 
the surface gravity: old stars are more “depleted“ in refractory elements (lower refractory-to-volatile ratios) 
than their younger counterparts. 

For FGK dwarf stars in the main sequence one does not expect significant changes in their atmospheric chemical abundances with age. This means
that the observed correlation between the T$_{c}$ slope and age probably reflects the chemical evolution in the Galaxy. 
We note that this is the simplest assumption we can make based on our limited current knowledge of stellar evolution, 
and we caution the reader that might be other effects that could severely affect the composition of stars  as a function of age.
For example, in the solar wind, elements with low first ionization potentials (FIP) are about four times more abundant than in the 
photosphere of the sun \citep[e.g.,][]{Geiss-98, Raymond-99}. However, this so-called FIP-effect, which will depend on age and 
has significant effect on coronal abundances \citep{Wood-06}, is not expected to affect the photospheric abundances.

In this context it is worth mentioning the work of \cite{Melendez-09}, where the authors discussed several possible explanations
of the solar ``peculiar'' abundances when compared to the solar twins. Most of the discussed possible effects (e.g., 
supernova pollution, early dust separation, etc.) however are not expected to have a dependence on age and can indeed be 
responsible for peculiarities in chemical composition for individual cases (stars).
In fact the median age of their comparison solar-twin stars was of 4.1 Gyr, a very similar to that of the Sun.
The authors also discussed the Galactic evolution effects, taking into account the possibility that the Sun may have
migrated from an inner Galactic orbit \citep{Wielen-96}, but ended up not considering it as a probable explanation.

\subsection{ T$_{c}$ slope and Galactic orbital parameters}

Several studies have shown that the mean of the apo- and pericentric distances (R$_{mean}$) are good indicators of the stellar birthplace 
\citep[e.g.,][]{Grenon-87, Edvardsson-93, Nordstrom-99,
Rocha-Pinto-04, Bensby-13}. \cite{Haywood-08a} shows that orbital parameters (R$_{mean}$) of the metal-poor and metal-rich thin disk stars are
significantly different from those of the main thin-disk population, and an outer and inner galactic disk origin was suggested for them
both, respectively.
A word of caution should be added here, if the radial migration proposed by \citet{Sellwood-02}%
\footnote{The term ``churning'' was introduced in \cite{Schonrich-09} for this type of migration.%
} is efficient, then using R$_{mean}$ as a proxy for the birthplace of a star could be dubious. However, \cite{Haywood-13} cast doubt upon
the efficiency of churning for contaminating the solar neighborhood.

\begin{figure}
\begin{center}
\begin{tabular}{c}
\includegraphics[angle=0,width=0.8\linewidth]{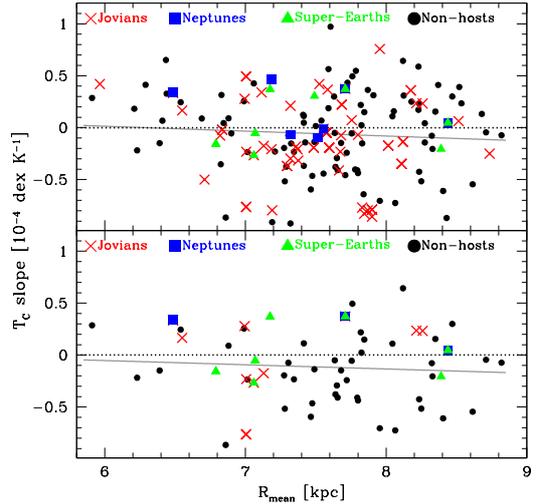}
\end{tabular}
\end{center}
\vspace{-0.6cm}
\caption{T$_{c}$ slopes versus R$_{mean}$ for the full sample (\textit{top}) and for the solar analogs (\textit{bottom}).
Gray solid lines provide linear fits to the data points.}
\label{fig_slope_rmean}
\end{figure}

In Fig.~\ref{fig_slope_rmean} we study the dependence of the T$_{c}$ slopes on R$_{mean}$, which we use as a proxy of the birth radii. 
Even if the correlation between the T$_{c}$ slopes and R$_{mean}$ is not strong and is not significant (at 1-$\sigma$ level), one can see
that most of the stars with the lowest mean galactocentric distances have steeper average T$_{c}$ slopes ($\approx0.071\pm0.065$) than the average of the stars 
with R$_{mean}$ = 8$\pm$1 kpc (T$_{c}$ $\approx-0.086\pm0.035$). The significance of this difference is 2.1$\sigma$ if one applies two sample \textit{t}-test.
For the solar analog sample, the significance of this difference is less significant (at a level of 1.5$\sigma$).

In Fig.~\ref{fig_slope_twins} we show the relation between the T$_{c}$ slopes and both the stellar age and the R$_{mean}$ for
solar twins in our sample. As one can see, all the correlations obtained for the full sample and sample of solar analogs are also valid for
the solar twins. Solar twins follow the general trends discussed in this section and in Sect. 3.1.

\cite{Rocha-Pinto-04}, among other authors, have already shown that R$_{mean}$ correlates with age%
\footnote{\cite{Wielen-96} have shown that this relation  can only be considered in a statistical sense.%
}in such a way that young objects (mostly younger than one Gyr)  all have R$_{mean}$ $\approx$ R$_{\odot}$, while older stars present a 
higher proportion of objects coming from different galactocentric radii.
In Fig.~\ref{fig_rmean_age} we plot R$_{mean}$ versus stellar age. One can note that there is a weak correlation between the two parameters
(with a correlation coefficient of $\approx$ 0.16 at a significance level of  $\approx$ 2$\sigma$),
even if not as an apparent way as in Fig. 7 of \cite{Rocha-Pinto-04}, probably because we do not have very young stars in our sample. 
However, in Fig.~\ref{fig_rmean_age} one can note that old stars (older than 6-7 Gyr) from the inner Galaxy ( R$_{mean}$ $<$ 7)
have predominantly positive T$_{c}$ slopes, while those in the solar circle have negative slopes.
In addition, the same figure shows that young stars with inner R$_{mean}$ values show an overabundance of
negative T$_{c}$ slopes, while their older counterparts mostly have positive slopes.

To test our findings observationally, i.e., if the T$_{c}$ slopes are different at different galactocentric distances (\textit{R}),
we used the Galactic abundance gradients derived in \cite{Lemasle-08, Lemasle-13}. 
Using the linear fits (provided by these authors)
of the gradient for each elements, we derived [X/Fe] abundance ratios at four galactocentric radii (6, 8, and 10 kpc). Then
we plotted [X/Fe] against T$_{c}$ (see Fig.~\ref{fig_lemasle}), and after performing a linear fit, we calculated the T$_{c}$ slopes. 
From the figure it is evident that at smaller galactocentric distances, the T$_{c}$ slope is steeper%
\footnote{We note that here one should changed the sign of the T$_{c}$ slopes to be consistent with the T$_{c}$ slopes derived in 
GH10,13.%
} which qualitatively (not quantitatively) agrees well with what we obtained in the T$_{c}$ versus R$_{mean}$ plot.
We note that in  \cite{Lemasle-08, Lemasle-13}, the authors used young Galactic Cepheids to derive the gradients. This means that
the gradients should be considered as current ones, which in turn means that they are not affected by any type of migration.
Although not quantitative, this comparison shows that at a fixed age (the present day) the 
T$_{c}$ correlates with the galactocentric radius. 
This is one more hint that R$_{mean}$ could be a relevant parameter for "single-age" populations 
(as already noted for the other end of the age domain explored in the paper, i.e. old stars, discussed in Fig.~\ref{fig_rmean_age}).

The tentative correlation between T$_{c}$ and the mean galactocentric distance of the stars, and the observational
results of \cite{Lemasle-08}, suggests that the chemical 
composition of the ``birth place'' is a factor partially responsible for the observed  T$_{c}$ trends.
The observed (weak) correlations suggest that at a fixed time the 
steeper  T$_{c}$ slopes (lower refractory-to-volatile ratios) are associated to stars that probably 
have originated in the inner Galaxy.  This result is in line with the recent chemical evolution models of the Milky Way invoking 
stellar migration \citep[e.g.,][]{Minchev-13}. Older stars are exposed to perturbations and scattering for a longer period of time
and have also had more time to migrate,
therefore some fraction of them originated in the inner Galaxy, while almost all the young stars have birth radii within the solar 
radius \citep[][]{Wang-13, Minchev-13}.

\subsection{ T$_{c}$ slope and planets}

In this letter we do not develop a long discussion about the signatures of planet formation on the abundance trends since it has already been
discussed  in detail by GH10,13. Using the same sample they have already shown that there is no statistically significant 
difference in T$_{c}$ slopes for planet-hosting stars (in particular for rocky planet hosts) and stars without any detected planetary companion.
In fact the average T$_{c}$ slope of the planet hosting stars (-0.08$\pm$0.38) is even slightly smaller (the difference is not significant)
than that of the non-host stars (-0.05$\pm$0.39) from the full sample.

Following our definition of solar analogs, we found that the average of the T$_{c}$ slope for planet hosting solar analogs is 
greater (0.012$\pm$0.31) than that of their non-host counterparts (-0.16$\pm$0.34). The Kolmogorov-Smirnov (K-S) statistics 
predict the $\approx$ 0.21  probability ($P_{KS}$) that these two 
subsamples came from the same underlying distribution for T$_{c}$ slope. At the same time, the same statistics predict a 
$P_{KS}$ $\approx$ 0.20 probability that they stem from the same underlying age distributions.
The latter can be seen in Fig.~\ref{fig_slope_age}: most of the planet-hosting 
stars tend to be relatively old ($\gtrsim$ 5 Gyr).
Moreover,  planet host and non-host samples show a different distribution of  R$_{mean}$ -- $P_{KS}$ $\approx$ 0.007.
As can also be seen in Fig. ~\ref{fig_slope_rmean}, 10 planet hosts out of 15 (66\%) have R$_{mean}$ smaller than 7.5 kpc (where
slopes are usually high) and only 16 out of 43 stars without detected planets (37\%) have similarly low R$_{mean}$ values.
Clearly the two subsamples are not consistent with respect to the mean galactocentric distance and age.
Interestingly, \cite{Haywood-09} has already shown that (giant) planet host stars tend to have smaller R$_{mean}$ and probably 
originate in the inner disk, which follow the same direction as our findings.

We note that the mean value of the T$_{c}$ slopes of the super-Earth hosts (M$_{planet}$ $<$ 0.05 M$_{J}$) is very close to the 
average value of the Jovian hosts (M$_{planet}$ $>$ 0.1 M$_{J}$), while the average T$_{c}$ slope of the three Neptunian hosts%
\footnote{The Neptunian hosts also host Jovian and super-Earth like planets.}%
(0.05 M$_{J}$ $\geq$  M$_{planet}$ $\geq$ 0.1 M$_{J}$) is greater. For a more detailed comparison  of the T$_{c}$ slopes for planet hosts
at different planetary mass regimes, we refer the reader to GH10,13.

These results suggest that the difference in T$_{c}$ slopes observed for solar analogs with and without planets is probably due to the 
differences in their ``birth places'' and times.

\section{Summary and conclusion}

We used a sample of 148 solar-type stars from GH10,13 to explore the main factors responsible for the abundance
trends with condensation temperature. For these stars the stellar atmospheric parameters and the T$_{c}$ slopes were taken from the 
above-mentioned studies, while the stellar ages,  Galactic orbital parameters, and velocity components are from \cite{Casagrande-11}.

Our study reveals a strong correlation between stellar ages and T$_{c}$ slope: old stars show steeper slope i.e., less 
refractory elements relative to volatiles. The same result reflects the statistically significant correlation between T$_{c}$ slopes
and surface gravity: more evolved (old) stars have a lower refractory-to-volatile ratio.

Moving one step further, we found tentative evidence that the T$_{c}$ slopes also correlate  with the mean galactocentric distance 
of the stars; this suggest that stars which probably have origin in the inner Galaxy (small R$_{mean}$) have steeper slopes.
The result fits well in the recent evolution picture of the Milky Way, showing that some fraction of old stars in 
the solar neighborhood might have its origin in the inner disk \citep[e.g.,][]{Minchev-13}.

Briefly exploring the possible reasons why one can see a difference in T$_{c}$ slopes for planet-hosting solar analogs and solar analogs
without detected planets, we found that in the current sample these two subsamples have different distributions of age and R$_{mean}$
that correlate with the T$_{c}$ slope. 
These differences might explain the differences in T$_{c}$ slope distribution, suggesting that there are no
signatures of planet formation in the observed abundance trends with the condensation temperature.

We may conclude that the T$_{c}$ slope depends on the age of a star at a fixed galactocentric radius.
At the same time, stars with smaller galactocentric radii show stepper  T$_{c}$ slopes at a fixed time (age).
In other words, the age and galactic birth place may determine the chemical pattern of the stars.

%
%________________________________________________________________
\begin{acknowledgements}

{We gratefully thank Misha Haywood and Ivan Minchev for fruitful comments and discussion. 
We gratefully acknowledge the (second) 
anonymous referee for a very exhaustive and honest review of the manuscript that led to many constructive comments and suggestions.
This work was supported by the European Research Council/European Community under the FP7 through Starting Grant agreement 
number 239953. V.Zh.A., S.G.S., and E.D.M are supported by grants SFRH/BPD/70574/2010, 
SFRH/BPD/47611/2008, and SFRH/BPD/76606/2011 from the FCT (Portugal), respectively.
NCS also acknowledges support in the form of a Investigador FCT contract funded by FCT/MCTES (Portugal) and POPH/FSE (EC).
G.I., S.B.L, and J.I.G.H. acknowledge financial support from the Spanish Ministry project MINECO AYA2011-29060, 
and J.I.G.H. also received support from the Spanish Ministry of Economy
and Competitiveness (MINECO) under the 2011 Severo Ochoa Program MINECO SEV-2011-0187.
PF is supported by the FCT and POPH/FSE (EC) through an Investigador FCT contract with application reference IF/01037/2013 and
and POPH/FSE (EC) by FEDER funding through the program "Programa Operacional de Factores de Competitividade - COMPETE.
}
\end{acknowledgements}
%________________________________________________________________

\bibliography{refbib}

\begin{thebibliography}{47}
\expandafter\ifx\csname natexlab\endcsname\relax\def\natexlab#1{#1}\fi

\bibitem[{{Adibekyan} {et~al.}(2012{\natexlab{a}}){Adibekyan}, {Delgado Mena},
  {Sousa}, {Santos}, {Israelian}, {Gonz{\'a}lez Hern{\'a}ndez}, {Mayor}, \&
  {Hakobyan}}]{Adibekyan-12a}
{Adibekyan}, V.~Z., {Delgado Mena}, E., {Sousa}, S.~G., {et~al.}
  2012{\natexlab{a}}, \aap, 547, A36

\bibitem[{{Adibekyan} {et~al.}(2013{\natexlab{a}}){Adibekyan}, {Figueira},
  {Santos}, {Hakobyan}, {Sousa}, {Pace}, {Delgado Mena}, {Robin}, {Israelian},
  \& {Gonz{\'a}lez Hern{\'a}ndez}}]{Adibekyan-13b}
{Adibekyan}, V.~Z., {Figueira}, P., {Santos}, N.~C., {et~al.}
  2013{\natexlab{a}}, \aap, 554, A44

\bibitem[{{Adibekyan} {et~al.}(2013{\natexlab{b}}){Adibekyan}, {Figueira},
  {Santos}, {Mortier}, {Mordasini}, {Delgado Mena}, {Sousa}, {Correia},
  {Israelian}, \& {Oshagh}}]{Adibekyan-13a}
{Adibekyan}, V.~Z., {Figueira}, P., {Santos}, N.~C., {et~al.}
  2013{\natexlab{b}}, \aap, 560, A51

\bibitem[{{Adibekyan} {et~al.}(2012{\natexlab{b}}){Adibekyan}, {Santos},
  {Sousa}, {Israelian}, {Delgado Mena}, {Gonz{\'a}lez Hern{\'a}ndez}, {Mayor},
  {Lovis}, \& {Udry}}]{Adibekyan-12b}
{Adibekyan}, V.~Z., {Santos}, N.~C., {Sousa}, S.~G., {et~al.}
  2012{\natexlab{b}}, \aap, 543, A89

\bibitem[{{Beaug{\'e}} \& {Nesvorn{\'y}}(2013)}]{Beauge-13}
{Beaug{\'e}}, C. \& {Nesvorn{\'y}}, D. 2013, \apj, 763, 12

\bibitem[{{Bensby} {et~al.}(2014){Bensby}, {Feltzing}, \& {Oey}}]{Bensby-13}
{Bensby}, T., {Feltzing}, S., \& {Oey}, M.~S. 2014, \aap, 562, A71

\bibitem[{{Casagrande} {et~al.}(2011){Casagrande}, {Sch{\"o}nrich}, {Asplund},
  {Cassisi}, {Ram{\'{\i}}rez}, {Mel{\'e}ndez}, {Bensby}, \&
  {Feltzing}}]{Casagrande-11}
{Casagrande}, L., {Sch{\"o}nrich}, R., {Asplund}, M., {et~al.} 2011, \aap, 530,
  A138

\bibitem[{{Dawson} \& {Murray-Clay}(2013)}]{Dawson-13}
{Dawson}, R.~I. \& {Murray-Clay}, R.~A. 2013, \apjl, 767, L24

\bibitem[{{Delgado Mena} {et~al.}(2014){Delgado Mena}, {Israelian},
  {Gonz{\'a}lez Hern{\'a}ndez}, {Sousa}, {Mortier}, {Santos}, {Adibekyan},
  {Fernandes}, {Rebolo}, {Udry}, \& {Mayor}}]{DelgadoMena-13}
{Delgado Mena}, E., {Israelian}, G., {Gonz{\'a}lez Hern{\'a}ndez}, J.~I.,
  {et~al.} 2014, \aap, 562, A92

\bibitem[{{Ecuvillon} {et~al.}(2006){Ecuvillon}, {Israelian}, {Santos},
  {Mayor}, \& {Gilli}}]{Ecuvillon-06}
{Ecuvillon}, A., {Israelian}, G., {Santos}, N.~C., {Mayor}, M., \& {Gilli}, G.
  2006, \aap, 449, 809

\bibitem[{{Edvardsson} {et~al.}(1993){Edvardsson}, {Andersen}, {Gustafsson},
  {Lambert}, {Nissen}, \& {Tomkin}}]{Edvardsson-93}
{Edvardsson}, B., {Andersen}, J., {Gustafsson}, B., {et~al.} 1993, \aap, 275,
  101

\bibitem[{{Figueira} {et~al.}(2013){Figueira}, {Santos}, {Pepe}, {Lovis}, \&
  {Nardetto}}]{Figueira-13}
{Figueira}, P., {Santos}, N.~C., {Pepe}, F., {Lovis}, C., \& {Nardetto}, N.
  2013, \aap, 557, A93

\bibitem[{{Fischer} \& {Valenti}(2005)}]{Fischer-05}
{Fischer}, D.~A. \& {Valenti}, J. 2005, \apj, 622, 1102

\bibitem[{{Geiss}(1998)}]{Geiss-98}
{Geiss}, J. 1998, \ssr, 85, 241

\bibitem[{{Gonzalez}(1997)}]{Gonzalez-97}
{Gonzalez}, G. 1997, \mnras, 285, 403

\bibitem[{{Gonzalez}(2011)}]{Gonzalez-11}
{Gonzalez}, G. 2011, \mnras, 416, L80

\bibitem[{{Gonzalez} {et~al.}(2010){Gonzalez}, {Carlson}, \&
  {Tobin}}]{Gonzalez-10}
{Gonzalez}, G., {Carlson}, M.~K., \& {Tobin}, R.~W. 2010, \mnras, 407, 314

\bibitem[{{Gonz{\'a}lez Hern{\'a}ndez} {et~al.}(2013){Gonz{\'a}lez
  Hern{\'a}ndez}, {Delgado-Mena}, {Sousa}, {Israelian}, {Santos}, {Adibekyan},
  \& {Udry}}]{Jonay-13}
{Gonz{\'a}lez Hern{\'a}ndez}, J.~I., {Delgado-Mena}, E., {Sousa}, S.~G.,
  {et~al.} 2013, \aap, 552, A6

\bibitem[{{Gonz{\'a}lez Hern{\'a}ndez} {et~al.}(2010){Gonz{\'a}lez
  Hern{\'a}ndez}, {Israelian}, {Santos}, {Sousa}, {Delgado-Mena}, {Neves}, \&
  {Udry}}]{Jonay-10}
{Gonz{\'a}lez Hern{\'a}ndez}, J.~I., {Israelian}, G., {Santos}, N.~C., {et~al.}
  2010, \apj, 720, 1592

\bibitem[{{Gonz{\'a}lez Hern{\'a}ndez} {et~al.}(2011){Gonz{\'a}lez
  Hern{\'a}ndez}, {Israelian}, {Santos}, {Sousa}, {Delgado-Mena}, {Neves}, \&
  {Udry}}]{Jonay-11}
{Gonz{\'a}lez Hern{\'a}ndez}, J.~I., {Israelian}, G., {Santos}, N.~C., {et~al.}
  2011, in IAU Symposium, Vol. 276, IAU Symposium, ed. A.~{Sozzetti}, M.~G.
  {Lattanzi}, \& A.~P. {Boss}, 422--423

\bibitem[{{Grenon}(1987)}]{Grenon-87}
{Grenon}, M. 1987, Journal of Astrophysics and Astronomy, 8, 123

\bibitem[{{Haywood}(2008{\natexlab{a}})}]{Haywood-08b}
{Haywood}, M. 2008{\natexlab{a}}, \aap, 482, 673

\bibitem[{{Haywood}(2008{\natexlab{b}})}]{Haywood-08a}
{Haywood}, M. 2008{\natexlab{b}}, \mnras, 388, 1175

\bibitem[{{Haywood}(2009)}]{Haywood-09}
{Haywood}, M. 2009, \apjl, 698, L1

\bibitem[{{Haywood} {et~al.}(2013){Haywood}, {Di Matteo}, {Lehnert}, {Katz}, \&
  {G{\'o}mez}}]{Haywood-13}
{Haywood}, M., {Di Matteo}, P., {Lehnert}, M.~D., {Katz}, D., \& {G{\'o}mez},
  A. 2013, \aap, 560, A109

\bibitem[{{Israelian} {et~al.}(2009){Israelian}, {Delgado Mena}, {Santos},
  {Sousa}, {Mayor}, {Udry}, {Dom{\'{\i}}nguez Cerde{\~n}a}, {Rebolo}, \&
  {Randich}}]{Israelian-09}
{Israelian}, G., {Delgado Mena}, E., {Santos}, N.~C., {et~al.} 2009, \nat, 462,
  189

\bibitem[{{Lemasle} {et~al.}(2013){Lemasle}, {Fran{\c c}ois}, {Genovali},
  {Kovtyukh}, {Bono}, {Inno}, {Laney}, {Kaper}, {Bergemann}, {Fabrizio},
  {Matsunaga}, {Pedicelli}, {Primas}, \& {Romaniello}}]{Lemasle-13}
{Lemasle}, B., {Fran{\c c}ois}, P., {Genovali}, K., {et~al.} 2013, \aap, 558,
  A31

\bibitem[{{Lemasle} {et~al.}(2008){Lemasle}, {Fran{\c c}ois}, {Piersimoni},
  {Pedicelli}, {Bono}, {Laney}, {Primas}, \& {Romaniello}}]{Lemasle-08}
{Lemasle}, B., {Fran{\c c}ois}, P., {Piersimoni}, A., {et~al.} 2008, \aap, 490,
  613

\bibitem[{{Mel{\'e}ndez} {et~al.}(2009){Mel{\'e}ndez}, {Asplund}, {Gustafsson},
  \& {Yong}}]{Melendez-09}
{Mel{\'e}ndez}, J., {Asplund}, M., {Gustafsson}, B., \& {Yong}, D. 2009, \apjl,
  704, L66

\bibitem[{{Minchev} {et~al.}(2013){Minchev}, {Chiappini}, \&
  {Martig}}]{Minchev-13}
{Minchev}, I., {Chiappini}, C., \& {Martig}, M. 2013, \aap, 558, A9

\bibitem[{{Mordasini} {et~al.}(2009){Mordasini}, {Alibert}, {Benz}, \&
  {Naef}}]{Mordasini-09}
{Mordasini}, C., {Alibert}, Y., {Benz}, W., \& {Naef}, D. 2009, \aap, 501, 1161

\bibitem[{{Nordstr{\"o}m} {et~al.}(1999){Nordstr{\"o}m}, {Andersen}, {Olsen},
  {Fux}, {Mayor}, {Mowlavi}, \& {Pont}}]{Nordstrom-99}
{Nordstr{\"o}m}, B., {Andersen}, J., {Olsen}, E.~H., {et~al.} 1999, \apss, 265,
  235

\bibitem[{{Nordstr{\"o}m} {et~al.}(2004){Nordstr{\"o}m}, {Mayor}, {Andersen},
  {Holmberg}, {Pont}, {J{\o}rgensen}, {Olsen}, {Udry}, \&
  {Mowlavi}}]{Nordstrom-04}
{Nordstr{\"o}m}, B., {Mayor}, M., {Andersen}, J., {et~al.} 2004, \aap, 418, 989

\bibitem[{{Ram{\'{\i}}rez} {et~al.}(2009){Ram{\'{\i}}rez}, {Mel{\'e}ndez}, \&
  {Asplund}}]{Ramirez-09}
{Ram{\'{\i}}rez}, I., {Mel{\'e}ndez}, J., \& {Asplund}, M. 2009, \aap, 508, L17

\bibitem[{{Ram{\'{\i}}rez} {et~al.}(2014){Ram{\'{\i}}rez}, {Mel{\'e}ndez}, \&
  {Asplund}}]{Ramirez-14}
{Ram{\'{\i}}rez}, I., {Mel{\'e}ndez}, J., \& {Asplund}, M. 2014, \aap, 561, A7

\bibitem[{{Raymond}(1999)}]{Raymond-99}
{Raymond}, J.~C. 1999, \ssr, 87, 55

\bibitem[{{Rocha-Pinto} {et~al.}(2004){Rocha-Pinto}, {Flynn}, {Scalo},
  {H{\"a}nninen}, {Maciel}, \& {Hensler}}]{Rocha-Pinto-04}
{Rocha-Pinto}, H.~J., {Flynn}, C., {Scalo}, J., {et~al.} 2004, \aap, 423, 517

\bibitem[{{Santos} {et~al.}(2001){Santos}, {Israelian}, \& {Mayor}}]{Santos-01}
{Santos}, N.~C., {Israelian}, G., \& {Mayor}, M. 2001, \aap, 373, 1019

\bibitem[{{Santos} {et~al.}(2004){Santos}, {Israelian}, \& {Mayor}}]{Santos-04}
{Santos}, N.~C., {Israelian}, G., \& {Mayor}, M. 2004, \aap, 415, 1153

\bibitem[{{Sch{\"o}nrich} \& {Binney}(2009)}]{Schonrich-09}
{Sch{\"o}nrich}, R. \& {Binney}, J. 2009, \mnras, 399, 1145

\bibitem[{{Schuler} {et~al.}(2011){Schuler}, {Flateau}, {Cunha}, {King},
  {Ghezzi}, \& {Smith}}]{Schuler-11}
{Schuler}, S.~C., {Flateau}, D., {Cunha}, K., {et~al.} 2011, \apj, 732, 55

\bibitem[{{Sellwood} \& {Binney}(2002)}]{Sellwood-02}
{Sellwood}, J.~A. \& {Binney}, J.~J. 2002, \mnras, 336, 785

\bibitem[{{Smith} {et~al.}(2001){Smith}, {Cunha}, \& {Lazzaro}}]{Smith-01}
{Smith}, V.~V., {Cunha}, K., \& {Lazzaro}, D. 2001, \aj, 121, 3207

\bibitem[{{Sousa} {et~al.}(2011){Sousa}, {Santos}, {Israelian}, {Mayor}, \&
  {Udry}}]{Sousa-11}
{Sousa}, S.~G., {Santos}, N.~C., {Israelian}, G., {Mayor}, M., \& {Udry}, S.
  2011, \aap, 533, A141

\bibitem[{{Wang} \& {Zhao}(2013)}]{Wang-13}
{Wang}, Y. \& {Zhao}, G. 2013, \apj, 769, 4

\bibitem[{{Wielen} {et~al.}(1996){Wielen}, {Fuchs}, \& {Dettbarn}}]{Wielen-96}
{Wielen}, R., {Fuchs}, B., \& {Dettbarn}, C. 1996, \aap, 314, 438

\bibitem[{{Wood} \& {Linsky}(2006)}]{Wood-06}
{Wood}, B.~E. \& {Linsky}, J.~L. 2006, \apj, 643, 444

\end{thebibliography}

\Online

\begin{appendix}

\section{}

\begin{figure}
\begin{center}
\begin{tabular}{c}
\includegraphics[angle=0,width=1\linewidth]{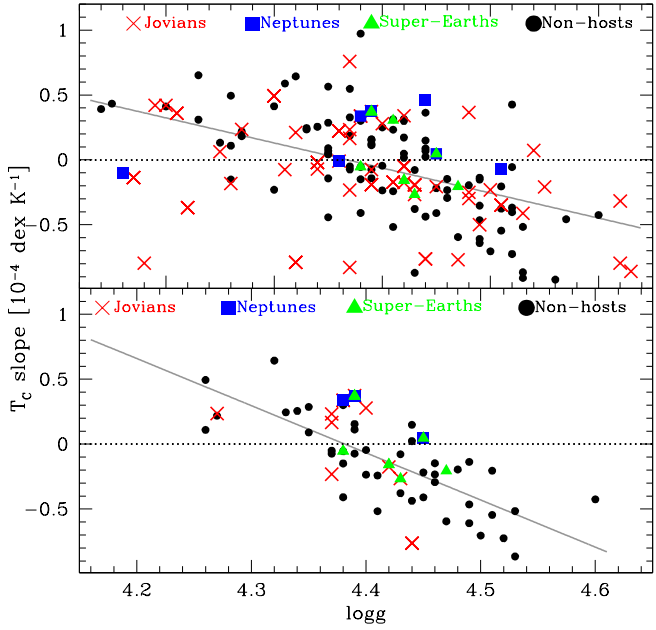}
\end{tabular}
\end{center}
\vspace{-0.6cm}
\caption{T$_{c}$ slopes versus surface gravity of the stars from the full sample (\textit{top}) and 58 solar analogs (\textit{bottom}).
Gray solid lines provide linear fits to the data points.}
\label{fig_slope_logg}
\end{figure}

\begin{figure*}
\begin{center}
\begin{tabular}{cc}
\includegraphics[angle=270,width=0.5\linewidth]{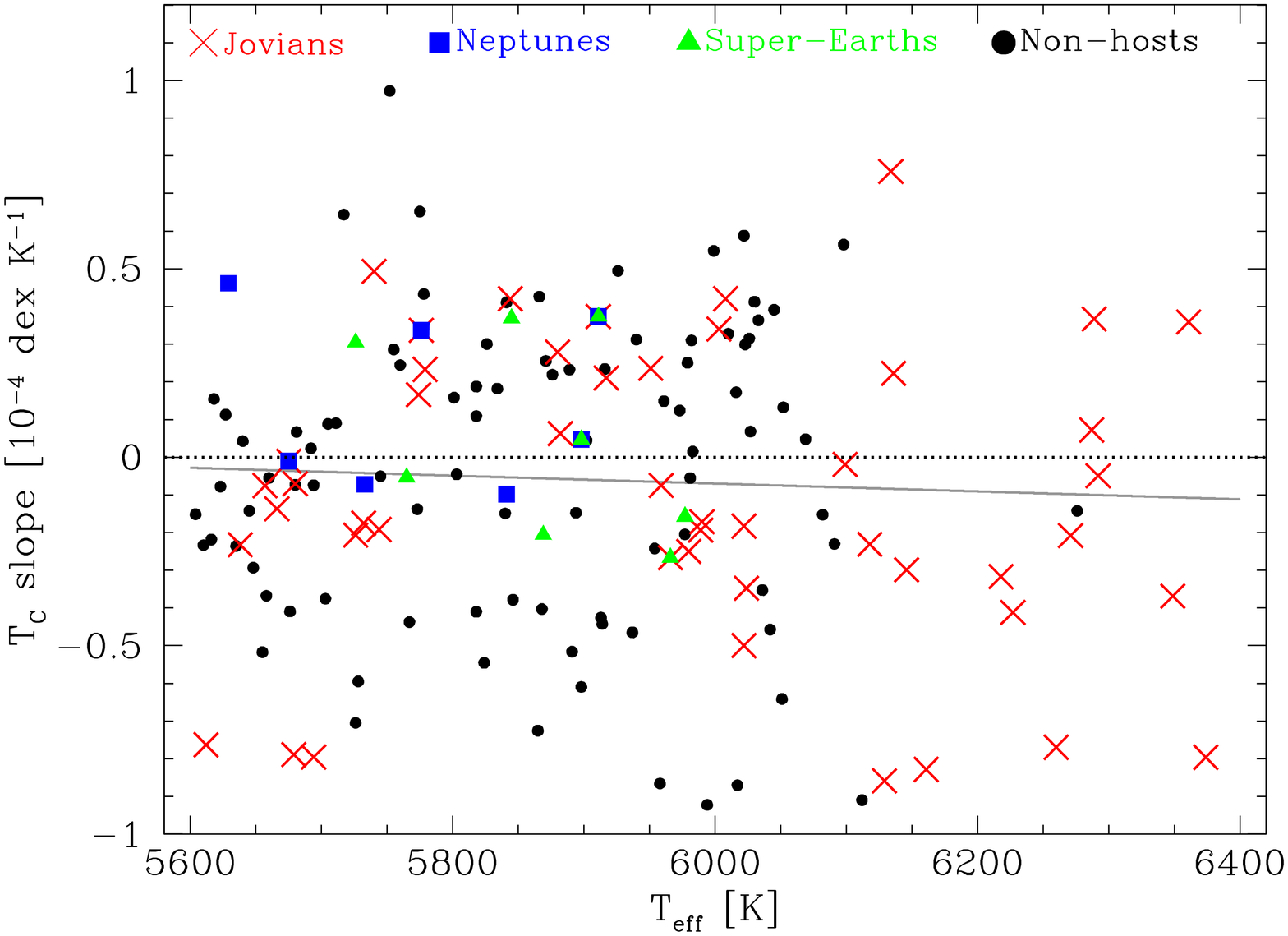}
\includegraphics[angle=270,width=0.5\linewidth]{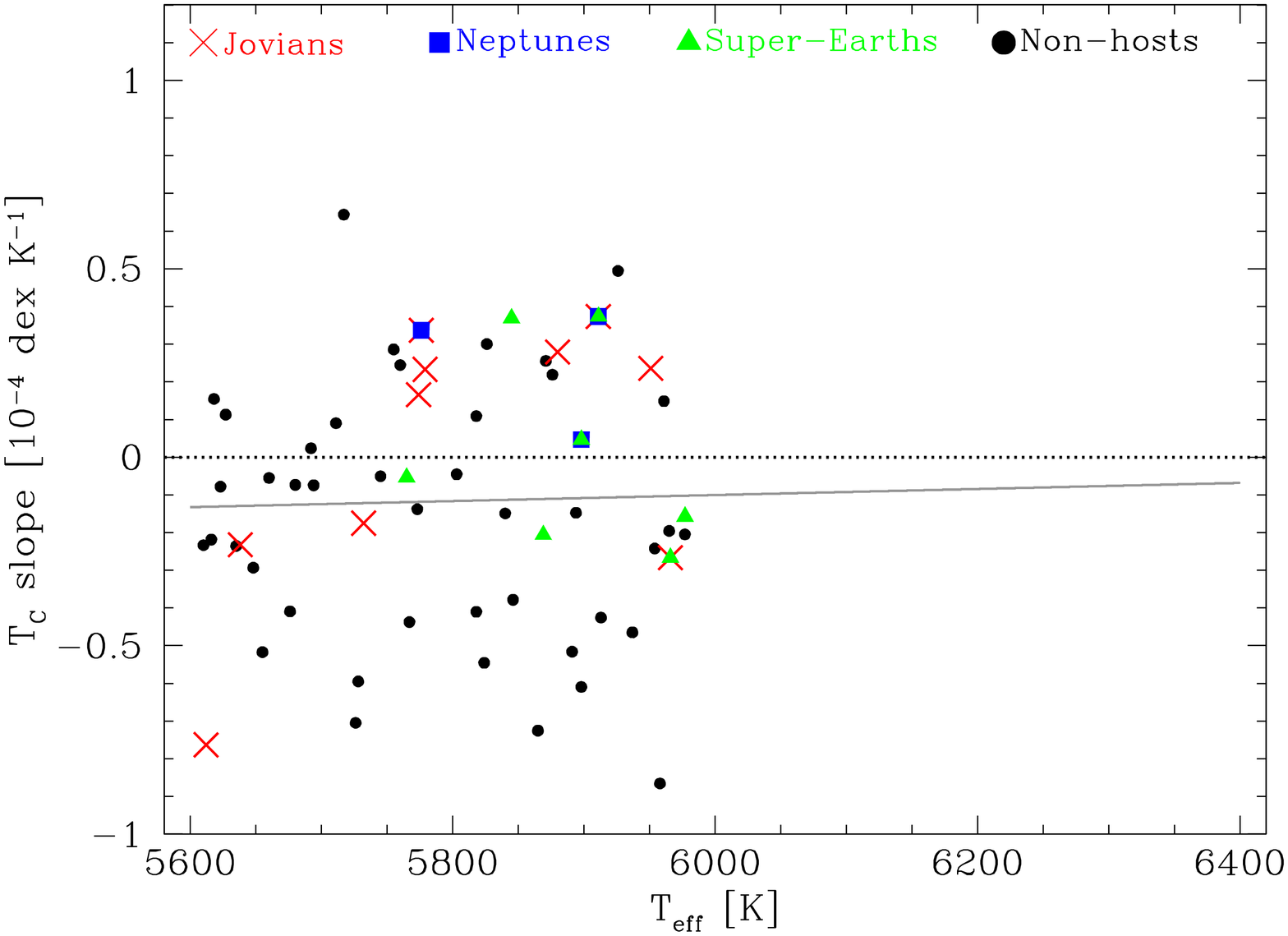}\tabularnewline[-0.05\columnwidth]
\includegraphics[angle=270,width=0.5\linewidth]{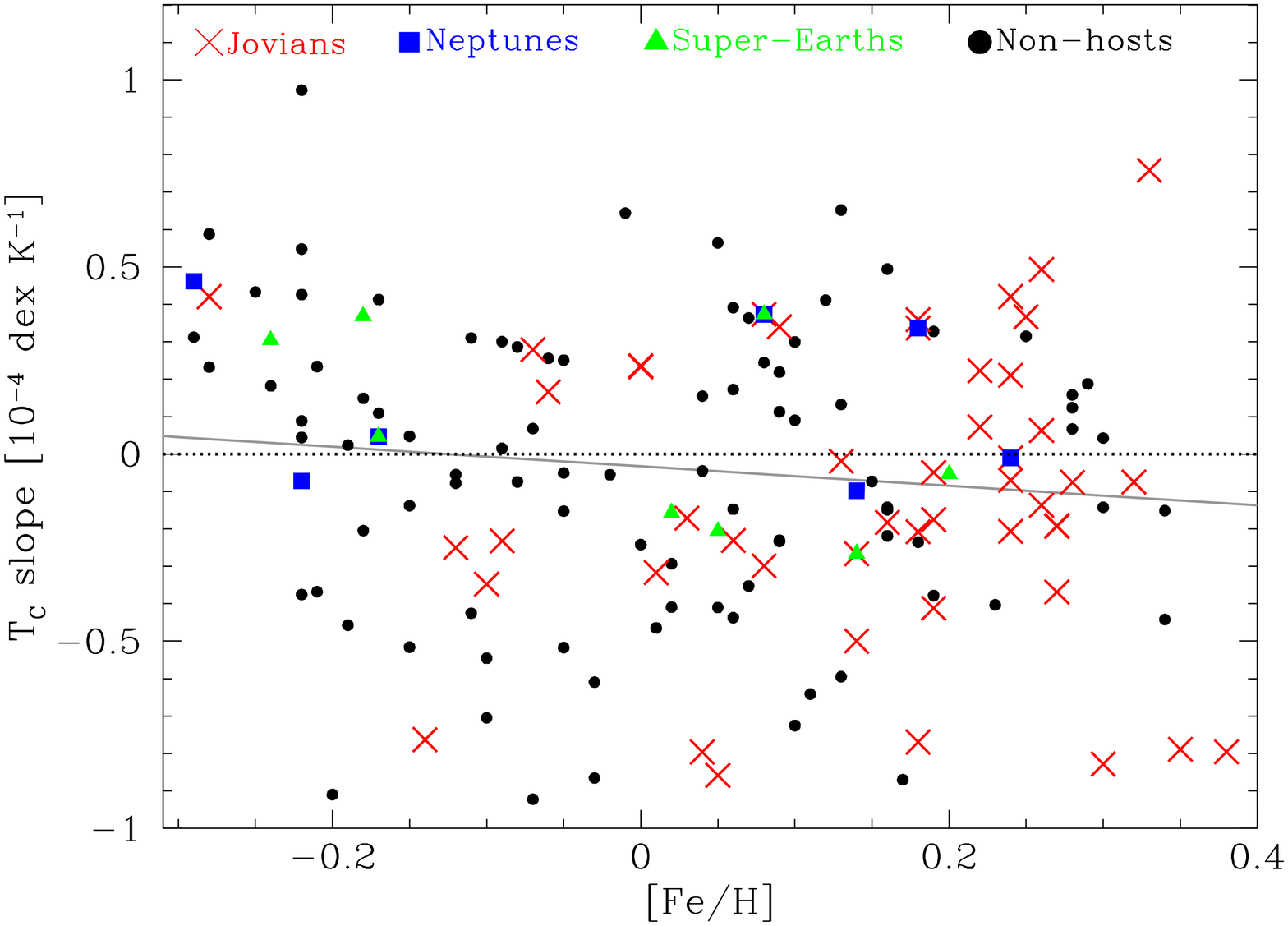}
\includegraphics[angle=270,width=0.5\linewidth]{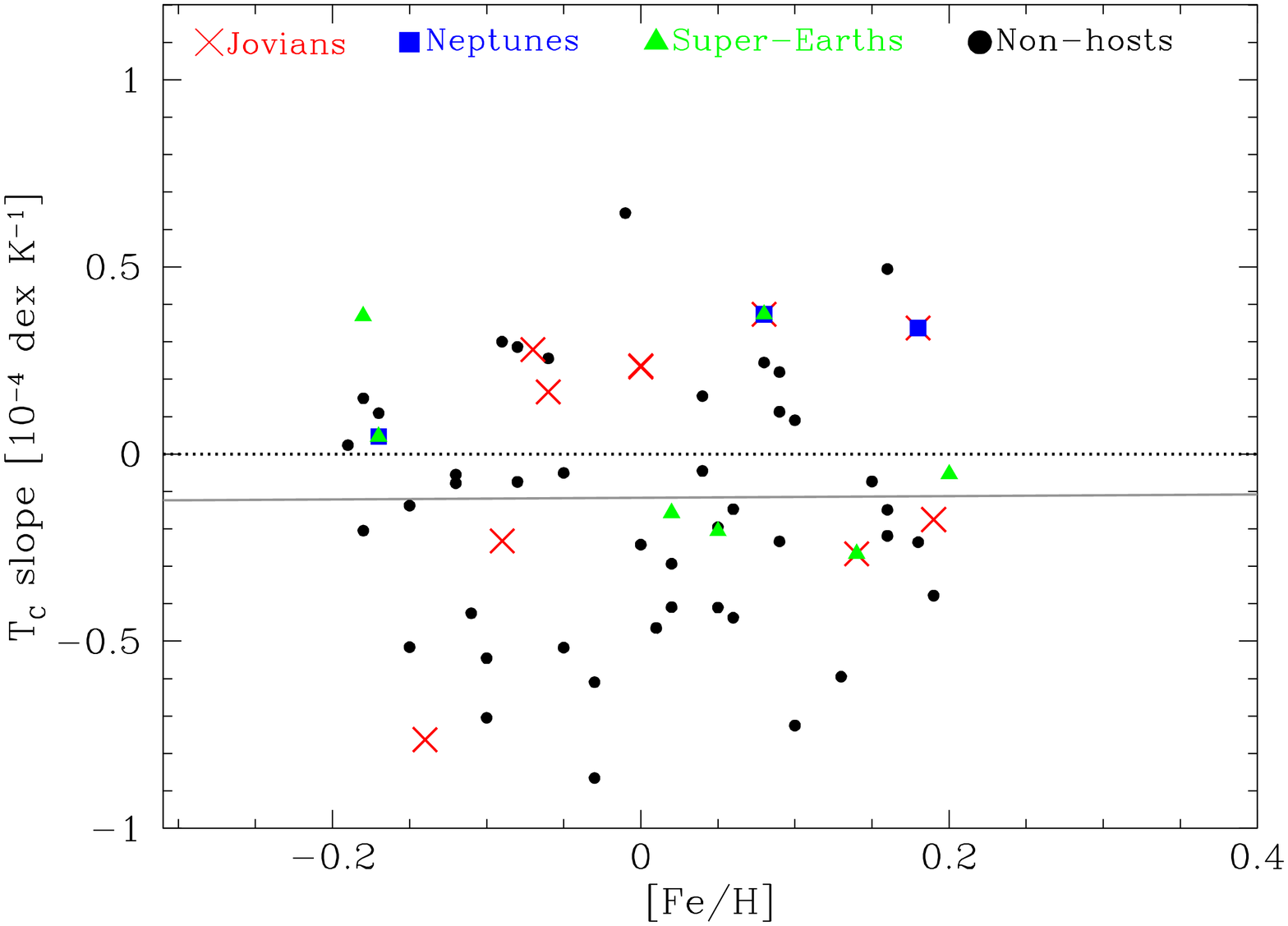}
\end{tabular}
\end{center}
\vspace{-0.5cm}
\caption{T$_{c}$ slopes versus atmospheric parameters of the stars from the full sample (\textit{left}) and solar analogs (\textit{right}).}
\label{fig_slope_atmos_param}
\end{figure*}

\begin{figure*}
\begin{center}
\begin{tabular}{cc}
\includegraphics[angle=270,width=0.5\linewidth]{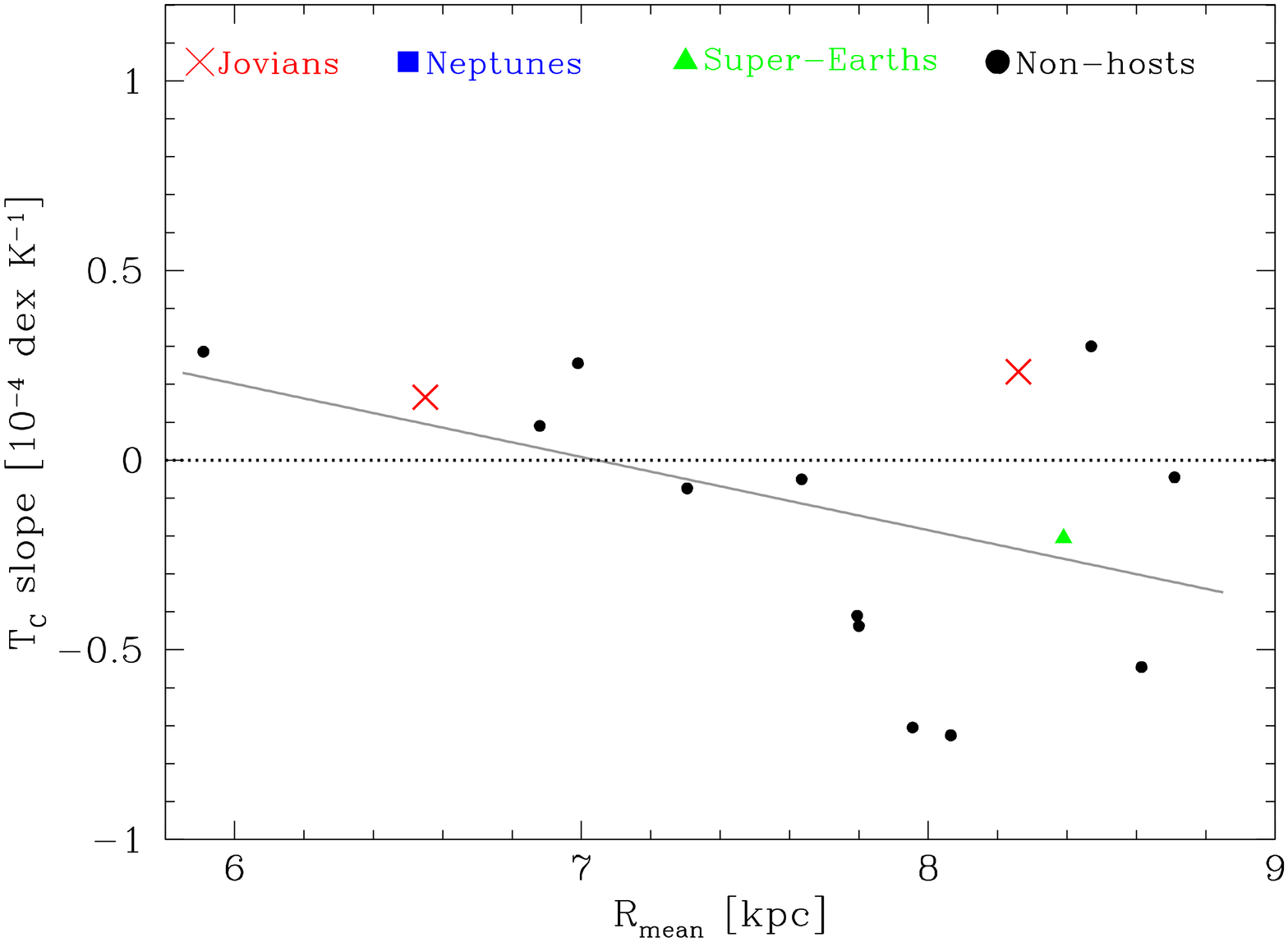}
\includegraphics[angle=270,width=0.5\linewidth]{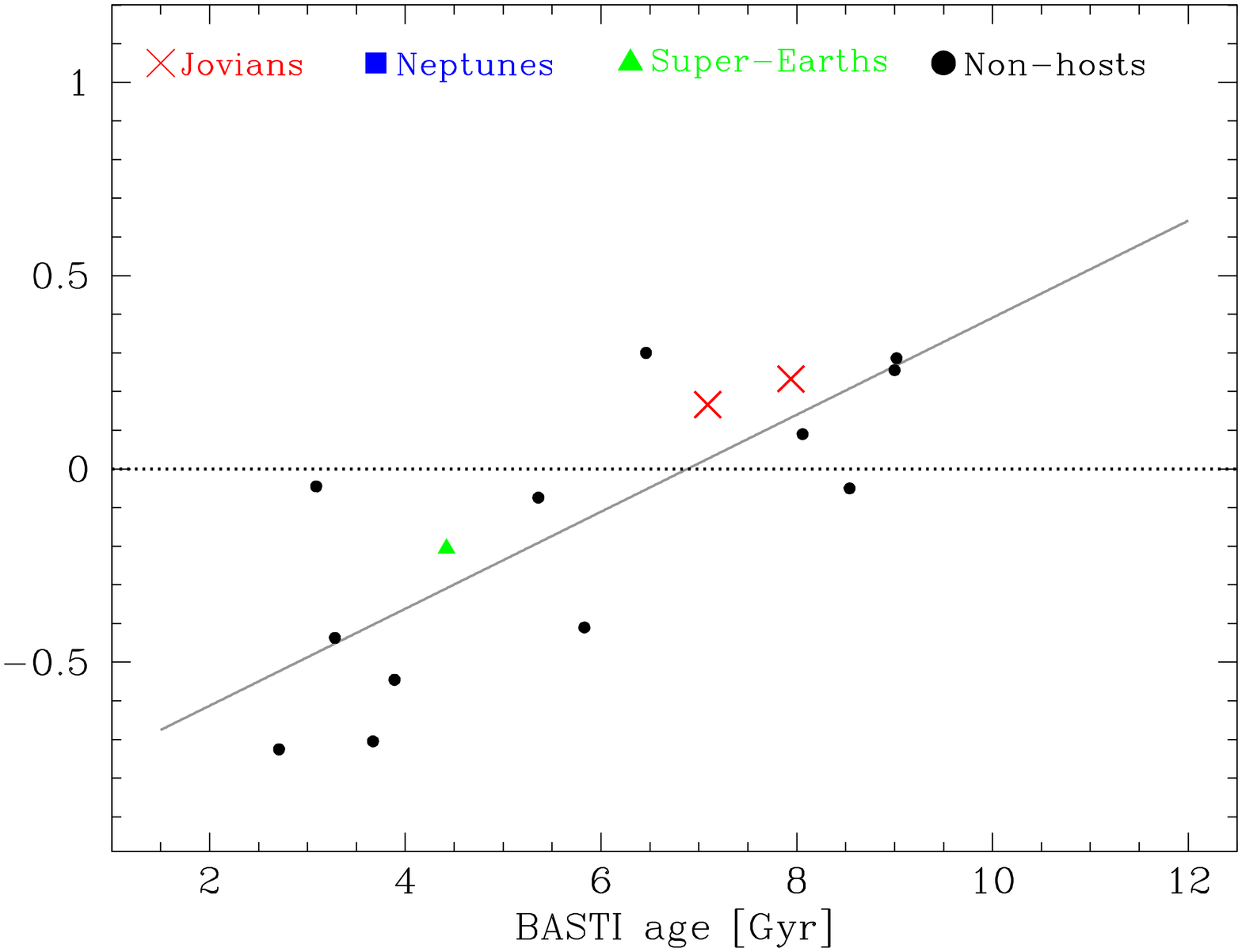}
\end{tabular}
\end{center}
\vspace{-0.5cm}
\caption{T$_{c}$ slopes versus age and the mean of the apo- and pericentric distances for the solar twins in our sample.
Gray solid lines provide linear fits to the data points.}
\label{fig_slope_twins}
\end{figure*}

\begin{figure}
\begin{center}
\includegraphics[angle=270,width=1\linewidth]{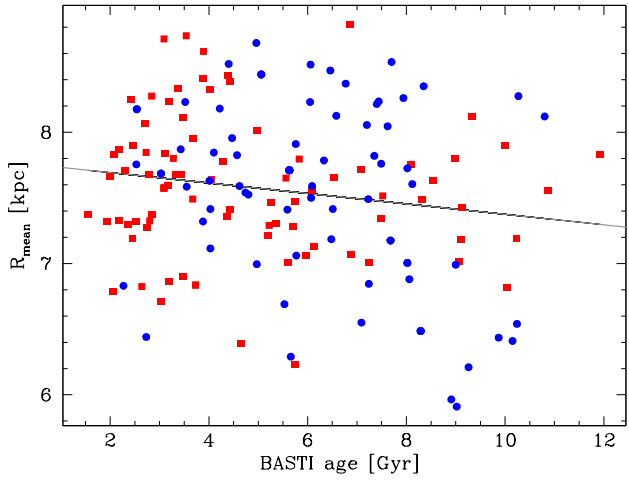}
\end{center}
\vspace{-0.5cm}
\caption{R$_{mean}$ versus age for the full sample. Red squares represent the stars with negative T$_{c}$ slopes, and 
the stars with positive T$_{c}$ slopes are marked by blue circles. 
Gray solid line provides linear fits to the full data points.}
\label{fig_rmean_age}
\end{figure}

\begin{figure}
\begin{center}
\begin{tabular}{c}
\includegraphics[angle=270,width=1\linewidth]{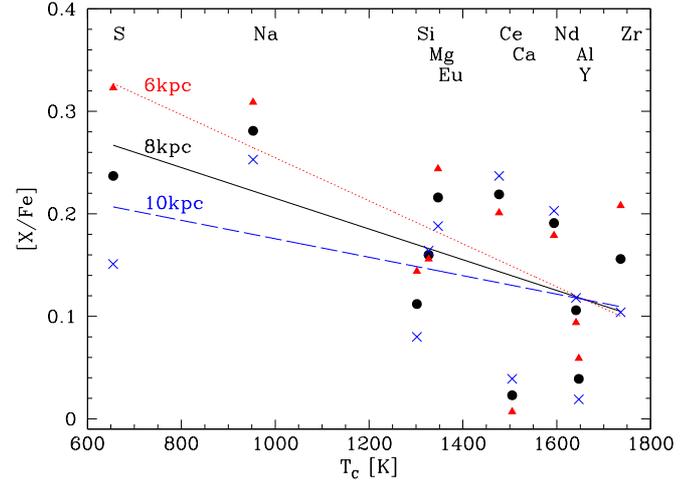}
\end{tabular}
\end{center}
\vspace{-0.6cm}
\caption{$[$X/Fe$]$ versus T$_{c}$ at three different galactocentric distances. The $[$X/Fe$]$ values derived using the
 Galactic abundance gradients of Galactic Cepheids from \cite{Lemasle-08, Lemasle-13}.
The  lines with different colors represent linear fits of the data.}
\label{fig_lemasle}
\end{figure}

\end{appendix}

\end{document}